\documentclass[aps,prb,preprint,groupedaddress]{revtex4}

\usepackage{graphicx}
\usepackage{textcomp}
\usepackage{gensymb}
\usepackage{amsmath}
\usepackage{array}

\makeatletter 
\def\tagform@#1{\maketag@@@{(\ignorespaces#1\unskip\@@italiccorr)}}
\makeatother

\RequirePackage[normalem]{ulem} 
\RequirePackage{color}\definecolor{RED}{rgb}{1,0,0}\definecolor{BLUE}{rgb}{0,0,1} 
\providecommand{\DIFaddbegin}{} 
\providecommand{\DIFaddend}{} 
\providecommand{\DIFdelbegin}{} 
\providecommand{\DIFdelend}{} 
\providecommand{\DIFaddbeginFL}{} 
\providecommand{\DIFaddendFL}{} 
\providecommand{\DIFdelbeginFL}{} 
\providecommand{\DIFdelendFL}{} 
\newcommand{\DIFscaledelfig}{0.5}
\RequirePackage{settobox} 
\RequirePackage{letltxmacro} 
\newsavebox{\DIFdelgraphicsbox} 
\newlength{\DIFdelgraphicswidth} 
\newlength{\DIFdelgraphicsheight} 
\LetLtxMacro{\DIFOincludegraphics}{\includegraphics} 
\newcommand{\DIFaddincludegraphics}[2][]{{\color{blue}\fbox{\DIFOincludegraphics[#1]{#2}}}} 
\newcommand{\DIFdelincludegraphics}[2][]{
\sbox{\DIFdelgraphicsbox}{\DIFOincludegraphics[#1]{#2}}
\settoboxwidth{\DIFdelgraphicswidth}{\DIFdelgraphicsbox} 
\settoboxtotalheight{\DIFdelgraphicsheight}{\DIFdelgraphicsbox} 
\scalebox{\DIFscaledelfig}{
\parbox[b]{\DIFdelgraphicswidth}{\usebox{\DIFdelgraphicsbox}\\[-\baselineskip] \rule{\DIFdelgraphicswidth}{0em}}\llap{\resizebox{\DIFdelgraphicswidth}{\DIFdelgraphicsheight}{
\setlength{\unitlength}{\DIFdelgraphicswidth}
\begin{picture}(1,1)
\thicklines\linethickness{2pt} 
{\color[rgb]{1,0,0}\put(0,0){\framebox(1,1){}}}
{\color[rgb]{1,0,0}\put(0,0){\line( 1,1){1}}}
{\color[rgb]{1,0,0}\put(0,1){\line(1,-1){1}}}
\end{picture}
}\hspace*{3pt}}} 
} 
\LetLtxMacro{\DIFOaddbegin}{\DIFaddbegin} 
\LetLtxMacro{\DIFOaddend}{\DIFaddend} 
\LetLtxMacro{\DIFOdelbegin}{\DIFdelbegin} 
\LetLtxMacro{\DIFOdelend}{\DIFdelend} 
\DeclareRobustCommand{\DIFaddbegin}{\DIFOaddbegin \let\includegraphics\DIFaddincludegraphics} 
\DeclareRobustCommand{\DIFaddend}{\DIFOaddend \let\includegraphics\DIFOincludegraphics} 
\DeclareRobustCommand{\DIFdelbegin}{\DIFOdelbegin \let\includegraphics\DIFdelincludegraphics} 
\DeclareRobustCommand{\DIFdelend}{\DIFOaddend \let\includegraphics\DIFOincludegraphics} 
\LetLtxMacro{\DIFOaddbeginFL}{\DIFaddbeginFL} 
\LetLtxMacro{\DIFOaddendFL}{\DIFaddendFL} 
\LetLtxMacro{\DIFOdelbeginFL}{\DIFdelbeginFL} 
\LetLtxMacro{\DIFOdelendFL}{\DIFdelendFL} 
\DeclareRobustCommand{\DIFaddbeginFL}{\DIFOaddbeginFL \let\includegraphics\DIFaddincludegraphics} 
\DeclareRobustCommand{\DIFaddendFL}{\DIFOaddendFL \let\includegraphics\DIFOincludegraphics} 
\DeclareRobustCommand{\DIFdelbeginFL}{\DIFOdelbeginFL \let\includegraphics\DIFdelincludegraphics} 
\DeclareRobustCommand{\DIFdelendFL}{\DIFOaddendFL \let\includegraphics\DIFOincludegraphics} 

\begin{document}
\title{Observation of enhanced coherence in Josephson SQUID cavities using a hybrid fabrication approach}

\author{S. Yanai}
\author{G. A. Steele}
\author{}
\author{}
\author{}
\author{}
\affiliation{
Kavli Institute of Nanoscience, Delft University of Technology, PO Box 5046, 2600 GA Delft, The Netherlands.}

\date{\today}

\begin{abstract}
We study the coherence of flux-tunable Josephson junction resonators made with two different fabrication processes. In the first process, devices are made using a single step of evaporation in which the resonator and the junctions of the SQUID are made at the same time. In the second process, devices are made with an identical geometry, but in which the resonators are made from a MoRe superconding layer to which an the junctions are added later in a second step. To characterize  the coherence of the two types of SQUID cavities, we observe and analyze the quality factor of their resonances as a function of flux and photon number. Despite a detailed cleaning process applied during fabrication, the single-step Al devices show significantly worse quality factor than the hybrid devices, and conclude that a the hybrid technique provides a much more reliable approach for fabricating high-Q flux-tunable resonators. 
\end{abstract}
\maketitle

\flushbottom
\maketitle
%
%
\thispagestyle{empty}

\section*{Introduction}

High coherence in superconducting resonant circuits is a highly desired property for any superconducting microwave component.  The reduction of microwave 
losses benefits many applications, including high kinetic inductance detectors\cite{hammer_ultra_2008,leduc_titanium_2010,szypryt_high_2016}, superconducting qubits\cite{bylander_dynamical_2011,paik_observation_2011}, Josephson parametric 
amplifiers\cite{bergeal_phase-preserving_2010,mutus_design_2013,zhong_squeezing_2013}, and superconducting hybrid systems
\cite{tabuchi_hybridizing_2014,yuan_large_2015,kubo_strong_2010}.
Recent research shows that the reduction of the participation ratio of the dielectric lossy layer with respect to the mode volume improves the coherence of superconducting qubits in 3D realization\cite{paik_observation_2011}.
It is also proven that this knowledge can be applied to two dimensional circuits, providing a higher scalability beneficial for quantum information processing\cite{bruno_reducing_2015}.

\begin{figure}
\centering
\includegraphics[width=80mm]{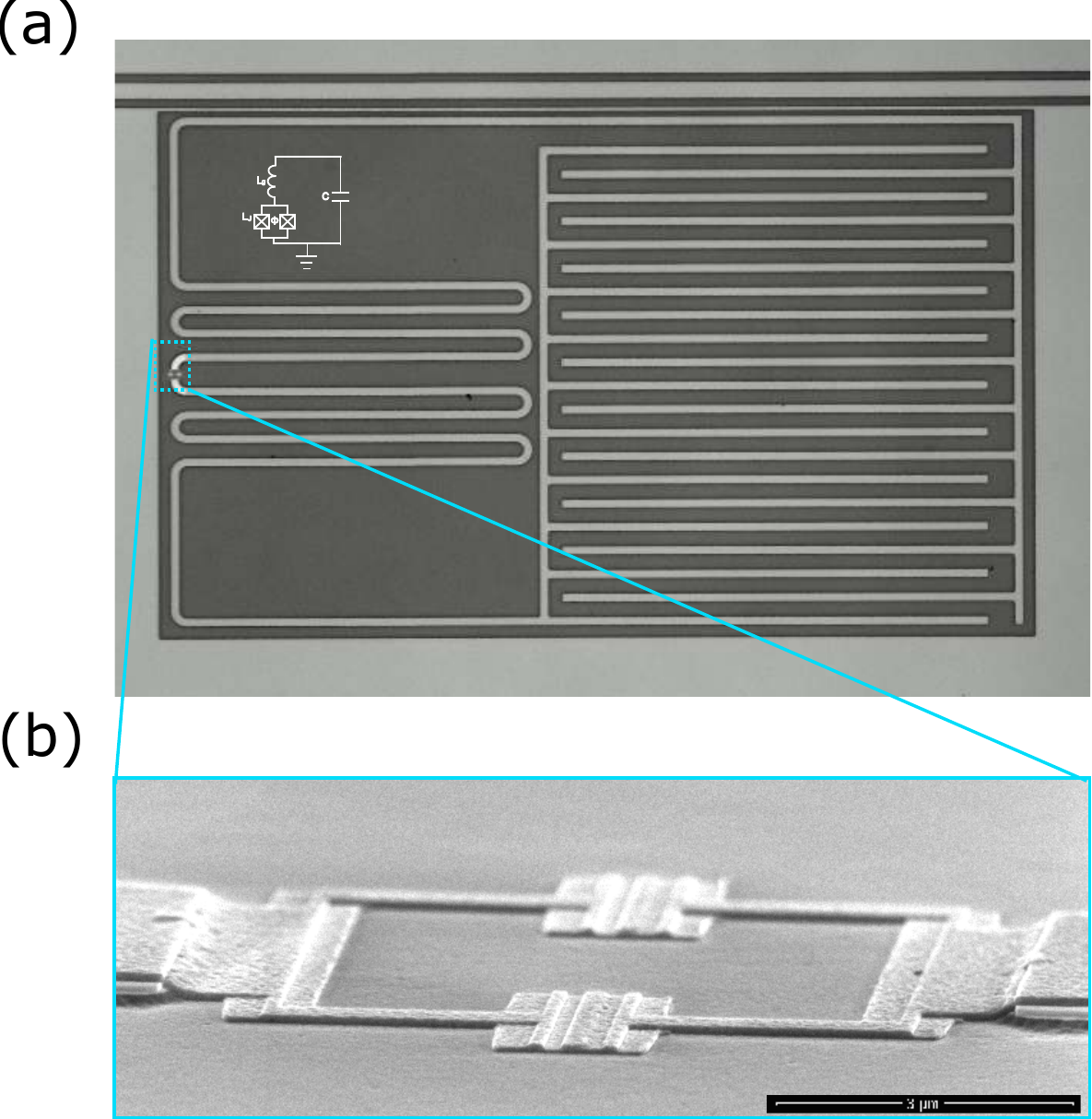}
\caption{(a) Optical microscopic image of the device.  In the cyan inset (dotted), the dc SQUID is galvanically coupled to the lumped element inductor. Circuit diagram is depicted on top left. (b) Scanning electron microscopic image of dc SQUID.}\label{fig6-1}
\end{figure}

Here, we use a study of the  internal quality  factor of flux-tunable resonators to explore the influence of the fabrication process on device coherence. Specifically,  we study identical device geometries fabricated with different techniques. The devices include lumped element resonators coupled to a microwave transmission line, along with the second device, which is a coplanar waveguide resonator terminated by a dc SQUID. In one of the fabrication processes, we made samples with a combination of reactive ion etching of MoRe lumped element resonators and lift-off of dc SQUIDs. The other procedure is a single lift-off process with a double angle evaporation of Al/AlOx/Al.
In this study, we find that the single-layer lift-off process resulted in internal quality factors two orders of magnitude lower than those yielded by the process with a reactive ion etching of resonators and lift-off of only the dc SQUIDs, suggesting the hybrid approach is superior for high-coherence superconducting circuits.

\section*{Results}
 Fig.~\ref{fig6-1}~(a) shows an optical image of one of the lumped-element resonators.  In the resonator, a dc SQUID is embedded inside the inductive part of the resonator, shown in Fig.~\ref{fig6-1}~(b). The dc SQUID provides a Josephson inductance that oscillates as a function of the externally applied flux with a periodicity of $\Phi_{0}=\frac{e}{2\hbar}$.  The total inductance of the resonator is given by $L_{g}+L_{J}(\Phi)$, where $L_{g}$ and $L_{J}(\Phi)$ are the geometric and Josephson inductance, respectively.

We have tested two different fabrication procedures for flux tunable resonators. In short, one of the types consist of a lumped element resonator made of 60~nm molybdeunum rhenium (MoRe) with standard Dolan bridge Al/AlOx/Al Josephson junctions.  The resonators are patterned with electron beam lithography (EBL) and reactive ion etching using gas of SF\textsubscript{6}+He. Subsequently, the dc SQUID with Al/AlOx/Al is evaporated using double angle evaporation. See methods section for full list of fabrication details.

\begin{figure}
\centering
\includegraphics[width=160mm]{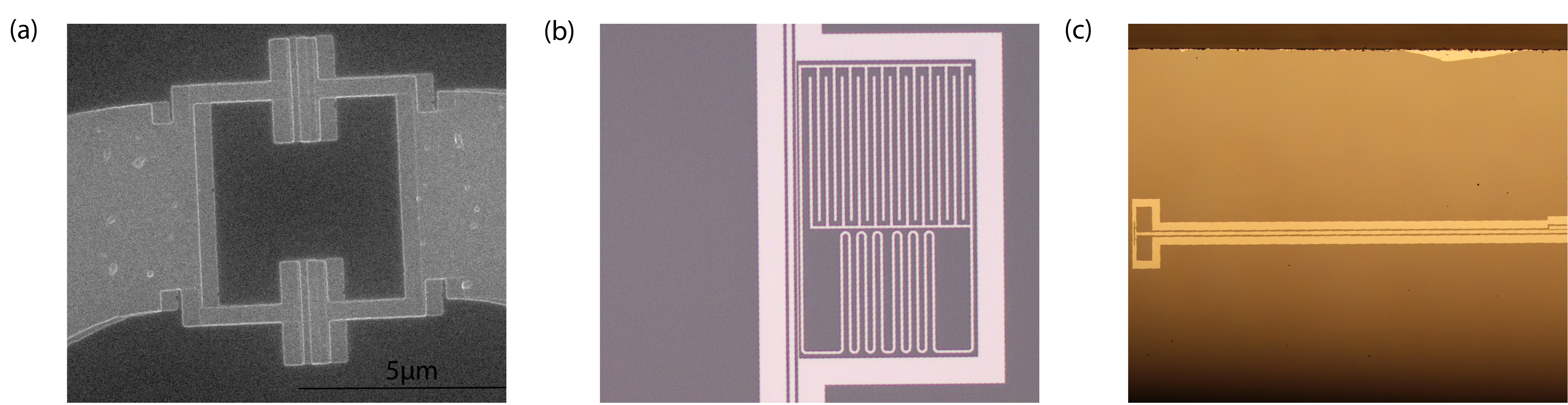}
\caption{(a) SEM image of the device fabricated with single lift-off. (b) Optical microscope image of the device. The picture does not contain a SQUID, but the measured device contains a dc SQUID the same as shown in (a). (c) CPW resonator terminated by a dc SQUID fabricated the same way as the device shown in Fig.~\ref{fig6-2}~(b).}\label{fig6-2}
\end{figure}

Fig.~\ref{fig6-2}~(a) shows an optical microscopic image of the device,  and Fig.~\ref{fig6-2}~(b) shows a high angle electron microscope image of the SQUID.  The SQUID loop is designed to be 5~$\times$~5~$\mu$m$^{2}$, but due to angle evaporation, the side is slightly less than 5~$\mu$m.  Each sample has multiple lumped element (LE) flux-tunable resonators, 
 which are side coupled in order to enable frequency multiplexing.  On each chip, we also include CPW and LE resonators without 
 SQUIDs as reference devices.   The other type of resonator was fabricated with a single step lift-off of the Al/AlO\textsubscript{x}/Al, including the microwave resonators and ground planes.  

The fabrication procedure of the flux-tunable resonator of this type is the same as the second half of the previous fabrication technique. The single lift-off 
procedure uses the same resist stack MMA/PMMA 950. The microwave resonator patterns including the Dolan bridges are exposed with EBL. The substrate is developed in a solution of MIBK: IPA 1:3 and IPA.  The substrate is de-scummed in oxygen plasma at 150 W for 30 s and cleaned in BHF for 30 s prior to shadow evaporation and the substrate is then lifted off in hot NMP.  Fig.~\ref{fig6-2}~(b) shows an image of a lumped element resonator fabricated with this technique. The structure of the resonator is the same, but due to the limited time of exposure, the ground plane is limited to 50 $\mu$m.  
We also tested a SQUID cavity made from a CPW shorted by a SQUID, as shown in Fig.~\ref{fig6-2}~(c). This device is configured in a reflection geometry.

All devices were mounted in a light-tight copper sample box and thermally anchored to the mixing chamber plate of a dilution refrigerator where the base temperature is below 20~mK. The microwave signal was attenuated at each stage inside the dilution refrigerator to provide thermalization of the microwave photons, and the signal was applied through a 50~$\Omega$ transmission line. The transmission or reflection spectrum was measured using a vector network analyser (VNA). The magnetic field was externally applied through the SQUID loop with a superconducting solenoid mounted on the mixing chamber of the dilution refrigerator.  No magnetic shielding was used during the measurements.

\begin{figure}
\centering
\includegraphics[width=160mm]{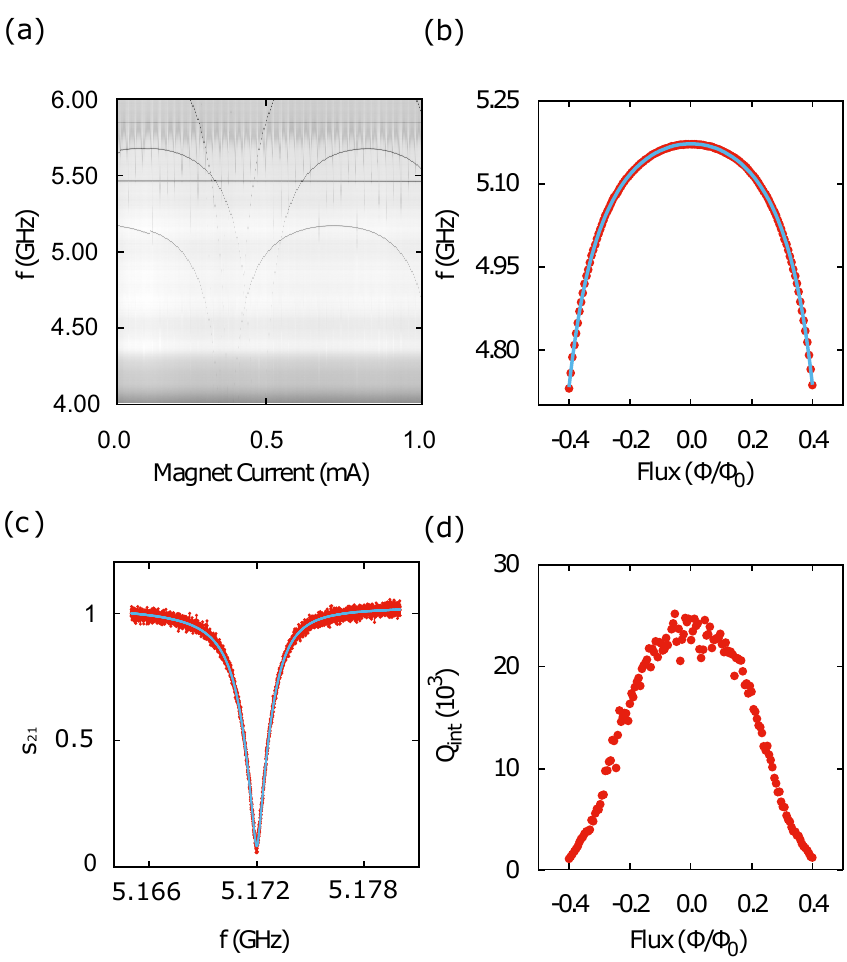}
\caption{Characterization of flux-tunable Josephson junction resonator, device \textbf{HY~1}. (a) Multiple resonances of different devices are measured simultaneously while applying external flux through the SQUID 
loops. Colour scale in the figure indicates depth in $|S_\mathrm{21}|$.  The $x$-axis represents current applied through the superconducting solenoid for approximately one and one-half full flux quantum.  The $y$-axis is the driving frequency $f$ with a VNA.  (b) Critical current of dc SQUID is estimated from the fitting to be 940~nA.  The $y$-axis of the figure represents the drive frequency $f$, and the red points indicate the resonance frequency, $f_{0}$, of one of the lowest modes in Figure~\ref{fig6-3}~a.
(c) Spectrum measurement of the resonator response, $S_\mathrm{21}$. The trace is taken at integer flux quantum, with an unloaded quality factor Q\textsubscript{int}
of 26294.306~\textpm~202.048. The normalisation procedure together with a Fano correction results in an unphysical S\textsubscript{21} greater than one (see SI for details). (d) Unloaded quality factor of the resonator as a function of flux through the SQUID loop.  }\label{fig6-3}
\end{figure}

In the measurements, the transmission spectrum $S_\mathrm{21}$ of the flux-tunable lumped element resonators was analysed.  Fig.~\ref{fig6-3} shows the basic 
characterization of the hybrid resonators. Fig.~\ref{fig6-3}~(a) shows a colour scale plot of $|S_\mathrm{21}|$ as a function of the flux and the flux dependence of resonance frequencies, $f_{0}$, of the devices.  The $y$ axis is the drive frequency $f$.  Each SQUID cavity on the chip was designed with a different geometric inductance, and therefore, at integer flux quantum it resonates at a different maximum frequency. This allows identifying each of the SQUID cavities in the multiple measurements. 

There are in all 5 resonant modes (Fig.~\ref{fig6-3}~(a)), where only 3 of them depend on the external magnetic field.  These have their integer flux quantum peaks around 5.2~GHz, 5.75~GHz, and 6.47~GHz.
It can be seen that the external magnetic fields corresponding to integer flux quantum for the three resonators are different, probably due to the flux induced in the SQUID loop during zero field cooling. The two modes independent of the external magnetic field correspond to the resonators that do not contain a dc SQUID. On the chip, there are three reference resonators for the purpose of characterizing the MoRe film.  One of the reference resonators is the lumped element linear resonator. The mode slightly below 5.5~GHz is the mode of the lumped element reference linear resonator.  The other two reference resonators have a CPW quarter wavelength geometry.  The mode around 5.85~GHz corresponds to the resonance frequency of one of the two CPW resonators.
The colour scale indicates the depth of the transmission spectrum.  As the response gets darker, the depth of the response gets bigger. The $x$-axis is the bias current applied to the superconducting solenoid.

The SQUID cavities are designedw with a relatively small Josephson inductance, $L_{J}$, whose critical current was designed on the order of $\mu A$ to accomodate reasonable number of intracavity photons in linear regime. To compensate the small L\textsubscript{J}, geometric inductance of the cavity was designed such that the modes of the SQUID cavities can be found in the measurement bandwidth without the effect of L\textsubscript{J}. 

Inductance and capacitance values were determined using  a combination of measurements of reference resonators and EM simulations. In order to determine the Josephson inductance from the participation ratio, we performed simulations in Sonnet to determine the effective geometric inductance seen at the position of the junction.  Due to the large kinetic inductance of MoRe film, the resonance frequency of the reference linear lumped element resonator deviates from the simulation by a large amount.  In the simulation, the circuit was designed with the same geometry as the reference lumped element linear resonator.  First, we adjusted the property of the metal in the simulation: we changed the value of the sheet inductance so that the resonance frequency in the simulation would match with the reference mode found in the measurement.  To find the Josephson inductance at integer flux quantum, we introduced an ideal lumped element inductive element into the circuit in the simulation. 
This ideal inductive element takes the role of a Josephson inductance in a real circuit and is adjusted such that the simulations match with the maximum value of the resonance frequency at integer flux quantum in the measurement. We found a total inductance of 2.9~nH and a Josephson inductance of the SQUID at integer flux quantum of 0.35~nH.  The participation ratio, including the geometric inductance and capacitance, can be further identified by looking at the frequency shift of the resonance upon increasing the ideal inductance. The shift arises to changes of the Josephson inductance due to flux in  the SQUID loop, consequently  changing the inductance participation ratio. The critical current can be estimated from the room temperature resistance $R_n$  over the SQUID and superconducting gap of 
the aluminium, which is discussed in supplementary material: we find agreement of these estimates based on $R_N$ to within about 50$\%$.

Fig.~\ref{fig6-3}~(c) shows the transmission spectrum of the resonant mode shown in Fig.~\ref{fig6-2}~(b) at zero flux quantum with an input power of -134.86~dBm at the device which corresponds to approximately one intracavity photon.  The response is not symmetric Lorentzian. Its asymmetry arises from a Fano resonance, the origin of which is most likely to be from an impedance mismatch of the transmission line including coaxial cables both outside and inside the fridge, and wire-bonds. We chose to normalize the response to the first point of the sweep.  The response is fitted to the skewed Lorentzian function with loaded and unloaded quality factors of 2150.253 and 26394.306~\textpm 202.048, respectively.

Fig.~\ref{fig6-3}~(d) shows the internal quality factor determined from the fits as a function of the flux through the SQUID loop.  The $y$-axis is the internal quality factor and the $x$-axis is the flux applied through the SQUID loop.  A reduction in internal $Q$ when approaching half-integer flux quantum is also observed in \cite{palacios-laloy_tunable_2008} and \cite{sandberg_tuning_2008}.  The origin of this away from integer flux quantum has not clearly understood.  \cite{palacios-laloy_tunable_2008} proposed that these losses could come from thermal fluctuations inside the cavity.  Sandberg et al.\cite{sandberg_tuning_2008} suggested that this loss is generated from subgap resistance.

Another possible source of the larger linewidth is flux noise.  The flux sensitivity of a SQUID is enhanced when approaching a half-integer flux quantum; the susceptibility to noise going through the SQUID loop is also increased, which leads to the reduction of quality as flux noise. The source of flux noise could be either extrinsic or intrinsic,  e.g. small fluctuations of the magnetic field on background which might exist constantly in the lab, and the quality decreases as the flux sensitivity is enhanced approaching half integer flux quantum, while the global magnetic field noise contributes to the loss.  Flux noise might also arise from origins intrinsic to the sample.  The model proposed for this phenomenon is a process of unpaired electrons occupancy defects and neighboring voids that causes fluctuations of the electron spin residing in a defect on the surface of the substrate or superconducting metal\cite{koch_model_2007}.   

\begin{figure}
\centering
\includegraphics[width=160mm]{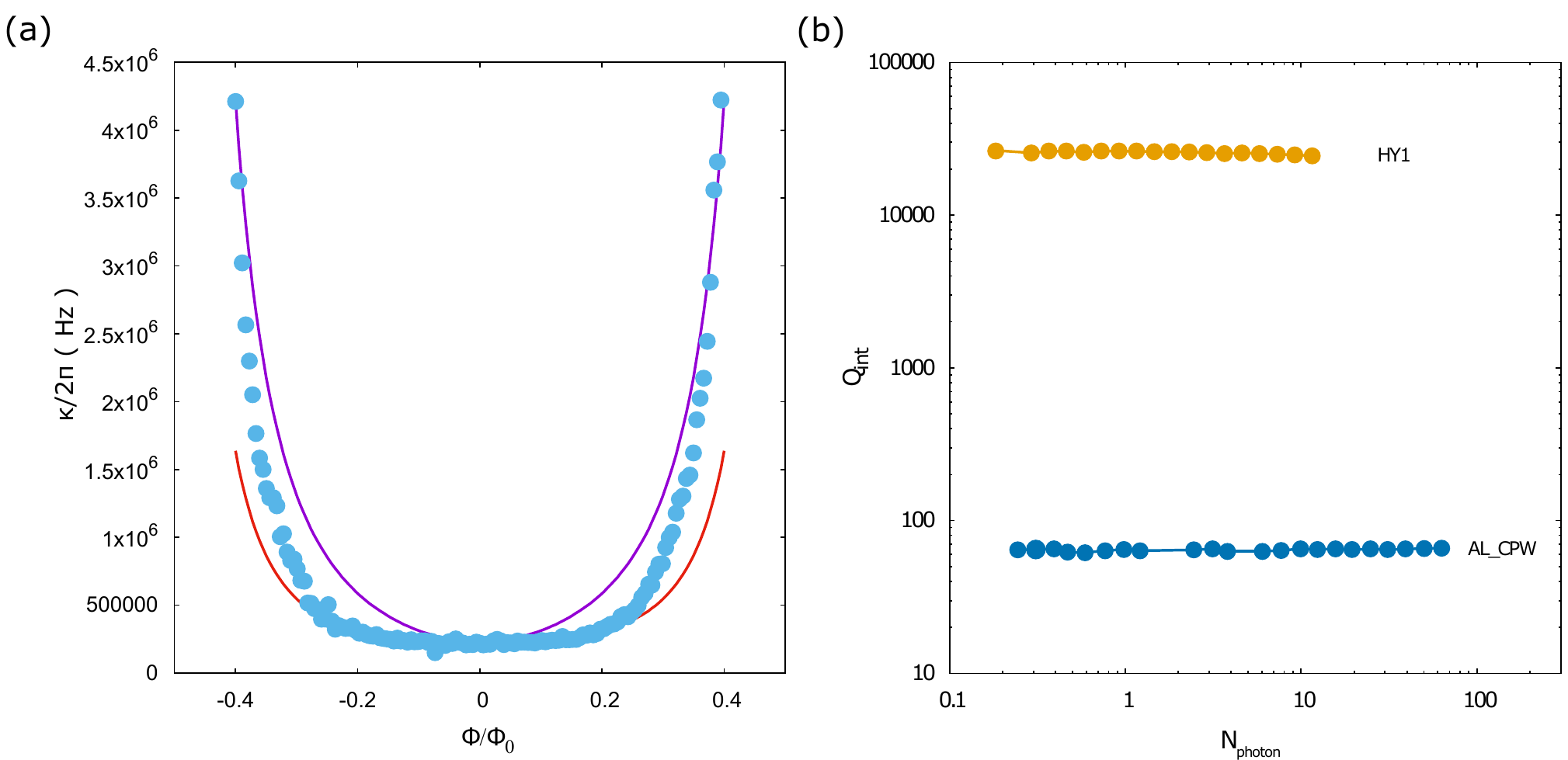}
\caption{(a) Internal loss rate  $\kappa_{int}$ vs flux of device HY~1. Cyan points are linewidths of unloaded resonator mode at different fluxes, $\omega_{0}/Q_{int}$. Purple and red solid lines are numerical simulations of $\kappa$, assuming the flux dependent loss is determined by a constant flux noise. The red line corresponds to the case when the flux noise is 1 m$\Phi_{0}$. The purple line corresponds to 2.6~m$\Phi_{0}$ of flux noise.
(b) Comparison of internal quality factors as functions of power. Yellow dots represents Hybrid flux tunable resonator (\textbf{HY~1}). Blue dots represent single step Al lift-off process (\textbf{AL\_CPW}). X-axis represents intra cavity photon numbers.}\label{fig6-4}
\end{figure}

In Fig.~\ref{fig6-4}, we study the flux dependence of the internal decay rates in more detail. We assume the leading term of the microwave loss away from integer flux quantum to be flux noise, and that the flux noise is constant as a function of the flux.  At that point, all the fluctuation noise of the flux contributes to microwawve loss via the SQUID. This gives the relation
\begin{equation}
\kappa_{int}=\sqrt{\kappa_{min}^{2}+\frac{d\omega}{d\Phi}\sigma_{\Phi}}\label{eq-11}
\end{equation}
where $\kappa_{min}=f(\Phi)/Q_{int}(\Phi)$ at integer flux quantum, and $\sigma_{\Phi}$ is the flux noise.
The cyan points in the figure are the experimental data which is defined by $f(\Phi)/Q_{int}(\Phi)$.  The red solid line is the plot with Eq.~\ref{eq-11} with 1~m$\Phi_{0}$ of flux noise. The purple solid line is the plot with 2.6~m$\Phi_{0}$ of flux noise.  The required flux noise need to explain our resonator linewidth was initially calculated from the change in the frequency under flux and the change in the internal quality factor. For a flux bias point of  -0.3~$\Phi_{0}$, we would estimate a flux noise of 1~m$\Phi_{0}$. With the 25 $\mu$m$^2$ area of our SQUID loop, this would correspond to stray noise field of 80 nT, often considered typical background magnetic field noise levels in a laboratory setting. This would suggest that the resisdual resonator linewidth of our hybrid devices could be limited by the absense of magnetic shielding the setup. 

Although the order of magnitude seems reasonable, the flux bias point dependence of the internal quality factor does not completely follow the model. In Fig.~\ref{fig6-4} we can see that neither of the estimated plots lies on top of the experimental data (cyan). The red line agrees near integer flux quantum and the purple plot matches around 0.4~$\Phi_{0}$. It is not entirely clear why the flux dependent internal losses do now follow the expectations from the model. One possibility is that the change of extracted internal quality factor with flux is not entirely due to flux noise. 

Fig.~\ref{fig6-4}(b) shows the power dependence of the internal quality factor of two devices.
Here, we have fixed the external flux such that the resonant mode is the maximum point, where $\pi\Phi/\Phi_{0}=0$. The measured transmission spectrum is a function of the power, and the unloaded quality factor is determined at different powers. The resonators are driven in the linear regime and fitted to a Lorentzian function as a function of power. The input power to the device has been varied from $\sim -142.86$~dBm to $\sim -112.86$~dBm, which is estimated by adding up the total attenuation of the input line (see supplementary material).  Fig.~\ref{fig6-4}(b) shows a comparison of two flux-tunable resonators. The orange points indicate the internal quality factors of one device at zero flux quantum fabricated with the hybrid process. The $x$-axis is the intracavity photon number, which is converted from total attenuation from the VNA to the sample. The blue points are the internal quality factors of a sample fabricated with the single lift off process in the CPW geometry.  This data is also taken at zero flux quantum.  The data is shown only for fits that show a good fit to the Lorentzian response, discarding higher powers where the response becomes non-linear.  The figure is cropped so that all the responses at the different powers fit well with a Lorentzian function. For the hybrid device, the dynamic range is limited around 10 intracavity photons.  From the figure, the internal quality factor of the hybrid device is much higher than the device made with the single lift off technique. 

Table~\ref{tb6-1} lists the results from different resonator tests. A first thing to note is the internal quality factors of reference resonators.  Two types of reference resonator, CPW and lumped element, are comparable to the values of hybrid1 and hybrid2.  This indicates that the internal quality factors of hybrid devices are not limited by dielectric loss from an interface between MoRe/AlO\textsubscript{x} or AlO\textsubscript{x} insulating layer of the junction.  However, the internal quality factors of the reference resonators are not as high as the ones reported previously for MoRe. For CPW resonators 
made on top of a sapphire substrate, the internal quality factor can reach 0.7~million at high power
\cite{singh_molybdenum-rhenium_2014}.  On a 
substrate of intrinsic silicon, the internal quality factor can be as high as 0.1 million\cite{singh_optomechanical_2014}.  Our 
device was fabricated 
on top of an intrinsic silicon substrate, and the MoRe was deposited with the same machine as for the devices in the references above. 

\begin{table}[]
\centering
\caption{Comparison of quality factors of the hybrid devices, the Al SQUID devices, and the MoRe reference resonators}
\label{tb6-1}
\begin{tabular}{|m{2.2cm}|m{3.2cm}|m{1.6cm}|m{1.6cm}|m{1.9cm}|m{3.4cm}|}
\hline
Device & Type & Q\textsubscript{int}(10\textsuperscript{3}) & Q\textsubscript{ext}(10\textsuperscript{3}) & f\textsubscript{0}& Metal\\
\hline
HY~1& hybrid & 25  & 2.3 & 5.172 GHz  & MoRe+Al/AlOx/Al \\ \hline
HY~2      & hybrid                 & 12         & 1.1        & 5.710 GHz  & MoRe+Al/AlOx/Al \\ \hline
AL\_LE~1      & Al SQUID lumped element  & 0.167      & 1.27       & 5.4786 GHz & Al/AlOx/Al      \\ \hline
AL\_LE~2      & Al SQUID lumped element  & 0.152      & 2.12       & 5.1054 GHz & Al/AlOx/Al      \\ \hline
AL\_CPW      & Al SQUID CPW             & 0.039      & 0.044      & 6.084 GHz  & Al/AlOx/Al      \\ \hline
REF\_LE~1      & lumped element reference & 11         & 1.73       & 5.44 GHz   & MoRe            \\ \hline
REF\_LE~2      & lumped element reference & 18         & 1.60      & 5.46 GHz   & MoRe            \\ \hline
REF\_CPW~1     & CPW reference            & 17         & 4.4        & 6.37 GHz   & MoRe            \\ \hline
REF\_CPW~2     & CPW reference           & 15         &7.72     & 5.83 GHz    & MoRe               \\ \hline
REF\_CPW~3     & CPW reference       &10            &7.4         &5.845 GHz   &MoRe              \\ \hline
\end{tabular}
\end{table}

\section*{Discussion}

We fabricated microwave flux-tunable resonators with a new recipe, in which the superconducting contacts are made between MoRe and AlOx.  We measured the coherence of these flux-tunable resonators and compared this with devices of the same design and a dc SQUID terminated CPW resonator (fabricated with the single step lift-off procedure).  The hybrid systems show high internal quality factors ($\sim 20000$ at zero flux quantum) two orders of magnitude larger than those of the devices made with the single step lift-off procedure.  The internal quality factors of hybrid devices are not limited by fabricating a Josephson junction inside the resonators.  

We suspect the observed low coherence in SQUID cavities of single lift-off devices is due to the combination of two factors: (1) Carbon contaminations on the S--M interface, as well as possible flip down of a resist wall after lift-off to metal electrodes. (2) A high participation ratio of the dielectric layers silicon--aluminium, aluminium--aluminium, and aluminium--air.  Further investigations might be beneficial, such as optimization of the oxygen plasma de-scumming, comparison of the internal quality factors of the single evaporation of Al with those from  shadow evaporated Al with the diffusive oxide layer in the middle.

As illustrated in Fig.~\ref{fig6-4}, the device fabricated with the hybrid process shows higher coherence than the CPW device fabricated with the single lift-off process. Also, the lumped element resonators fabricated with the single lift-off process show low internal quality factors compared with the hybrid devices, which can be seen in Table~\ref{tb6-1}. 

One possible cause of loss in the single lift-off process could insufficient surface treatment. Device patterned with the single lift-off process could contaminate the surface with carbon residual layer due to insufficient surface preparation, along with $\sim$~2~nm AlO\textsubscript{x} on top of the carbon layer, which could arise a product of a reaction of unpassivated Al with resist contamination or solvent from the development, or has resulted from heating the substrate\cite{quintana_characterization_2014}. In \cite{quintana_characterization_2014} it is shown that the estimated loss tangent of 2~nm of the resist residue left due to insufficient de-scumming could leads to $\mathrm{\delta_{TLS}\sim 3\times 10^{-3}}$. While large, it is unlikely that the contamination would contribute to the dielectric loss in the resonator with unity participation ratio. 

Our devices we have also observed contamination around the resonator electrode edges due to  what appears to from a collapsing side wall of the resist stack on top of the Al electrodes (see supplementary material). Although this has been discussed before in the field, and reported in theses \cite{slichter_quantum_2011}, it is not clear if this would present a significant enough loss to explain our observations. This sidewall contamination, though, can be mitigated by changing the bottom layer MMA into another resist layer, such as PMGI.  

Another possible source of dielectric loss could be the AlO\textsubscript{x} layer between the two aluminium layers from the shadow evaporation.   This insulating layer inside the SQUID cavity could form a  capacitance where energy is stored, and degrade the coherence of the mode.  This dielectric loss is not a leading cause of microwave energy loss in 3D realizations of a superconducting qubit \cite{paik_observation_2011}\cite{chu_suspending_2016}\cite{rigetti_superconducting_2012}, which are also made in a single-step shadow evaporation process. This could be due to the geometry of their 3D devices:  the dielectric loss resides in the substrate--metal, substrate--air, or metal--air interfaces, and possibly even in the AlO\textsubscript{x} layer there is a reduction of this loss to the cavity field by increasing the mode volume of the cavity where there is no dielectric loss in vacuum. 

Stray magnetic fields from the lack of shielding can also influence the internal quality factor of Aluminum films, although reports suggest that field up to 100 uT (4 times the field of the earth) can still result in quality factors in excess of 10$^4$ \cite{flanigan_magnetic_2016}

Although the exact origin of poor internal quality factor of the single layer devices studied here is not clear, the alternative technique presented here based on hybrid resonators clearly is a viable solution to achieving high internal quality factor in superconducting Josephson cavities. 

\section*{Methods}

The substrate, high resistivity ($\rho\sim10000~\Omega cm$) silicon (100) wafer, is first cleaned in an RCA 1 solution at {70}\textdegree C for 10 minutes, followed by piranha cleaning at {90}\textdegree C for 10 minutes.  The purpose of this cleaning is to remove particles and any organic chemicals on the surface of the substrate.  Subsequently, the substrate is cleaned in a BHF solution to remove native oxide and terminate the surface with hydrogen, which has a hydrophobic surface. 

MoRe superconducting metal is then deposited with rf sputtering with the thickness of 60~nm immediately after the BHF wet etching.  An S1813/Tungsten/PMMA tri-layer resist stack is then spin coated on the surface of the substrate and afterwards the microwave circuits are patterned with electron beam lithography (EBL) and reactive ion etching (RIE). Finally, the substrate
is developed in a solution ofMIBK: IPA 1:3, followed by IPA in order to stop the developing process.  The surface of the exposed areas is tungsten.  In the RIE, 
tungsten, S1813, and MoRe are etched sequentially with SF\textsubscript{6}/He, oxygen, and SF\textsubscript{6}/He, respectively.  After etching, the resist mask is removed in hot PRS 3000 resist stripper.  

Next, the substrate is spin coated with a bi-layer MMA /
PMMA 950 resist stack, and patterns for the Dolan bridges are exposed. The substrate is developed in a mixture of MIBK: IPA 1:3 and IPA to stop the development.  Right before the evaporation of the junctions, the substrate is de-scummed in the oxygen plasma
at 150 W for 30 s and then cleaned in BHF for 30 s. With this procedure we can remove all contaminants
and native oxide on the surface of the MoRe and therefore obtain a much better MoRe--Al interface.  Immediately after 
the BHF cleaning, the substrate is loaded in an electron beam evaporator, and Al is evaporated from {11}\degree from each side over the Dolan bridges with an oxidization step in the middle. 
The substrate is then lifted off with a hot NMP solution until the resist stack comes off completely.

\newpage

\bibliography{main}

%


%

%

%

%

\section*{Contributions}
S.Y. fabricated the devices and performed the measurements. S.Y. analyzed the data and wrote the manuscript with input from G.A.S. 

\section*{Competing interests}
The author(s) declear no competing interests. 
\newpage
\section*{\large Supplementary Material : Observation of enhanced coherence in Josephson SQUID cavities using a hybrid fabrication approach}
\section{Derivation of the flux-tuning curve of the SQUID cavity}

Here we start by reviewing an analytical expression for the Josephson inductance. Due to the large designed $L_{J}=300$~pH, we have neglected the geometric inductance of the SQUID loop $L_{g}^{SQ} \sim 10$~pH in the following analysis.  (A detailed analysis of such circuits can be found in \cite{Palacios_thesis}).  We consider a symmetric dc SQUID where the two Josephson junctions have the same critical current, $I_{1}$, $I_{2}$, where $I_{n} = I_{c0} sin \gamma _{n}$, with $n=1,2$, where $n$ indicates the two Josephson junctions of the SQUID.  The total current is given by $I=I_{1}+I_{2}$, and circulating current by $J=(I_{1}-I_{2})/2$.  These two equation can be expressed in the following forms:

\begin{equation}
I=2I_{c0}\mathrm{\cos}\phi \mathrm{\sin}\gamma\label{eq-1}
\end{equation}

\begin{equation}
J=I_{c0}\mathrm{\sin}\phi \mathrm{\sin}\gamma
\end{equation}\label{eq-2}
where $\phi=\pi\frac{\Phi}{\Phi_{0}}$ is the flux frustration and $\gamma=\frac{1}{2}(\gamma_{1}+\gamma_{2})$ is the phase difference across the SQUID, $\theta_{2}-\theta_{1}$, shown in  Fig.~\ref{FigS1}~(a).  An important consequence of these expressions is the appearance of an inductance $L_J$ associated with the Josephson junction in the form
\begin{equation}
V=L_{J}(\phi)\frac{dI}{dt}\label{eq-3}
\end{equation}
From the Josephson relation, we can obtain

\begin{equation}
L_{J}=\varphi_{0} \frac{d\gamma/dt}{dI/dt}\label{eq-4}
\end{equation}
where $\varphi_{0} = \frac{\Phi_{0}}{2\pi}$ is the reduced flux quantum. 

In the following, we will neglect $d{\phi}/dt$.  This is justified due to the small geometric inductance of the SQUID loop.  With $d{\phi}/dt$=0, we have the following for $dI/dt$ from Eq.~\ref{eq-1}.
\begin{equation}
\frac{dI}{dt}=I_{c}\cos{\phi}\cos{\gamma}\frac{d\gamma}{dt},\label{eq-5}
\end{equation}
Here, $I_{c}$ is the critical current of the SQUID.  Using Eq.~\ref{eq-4}, the Josephson inductance is given by 
\begin{equation}
L_{J}=\varphi_{0}\frac{1}{I_{c}\cos{\phi}\cos{\gamma}}\label{eq-6}.
\end{equation}
For small excitation powers, $\mathrm{\cos}\gamma\sim 1-\frac{\gamma^{2}}{2}$, using 
$1/{(1-\gamma^{2}/2)}\sim 1+\gamma^{2}/2$, which gives 
\begin{equation}
L_{J}=\varphi_{0}\frac{1}{I_{c}\cos{\phi}}(1+\frac{\gamma^{2}}{2})\label{eq-7}.
\end{equation}
Now we consider only the linear term in Eq.~\ref{eq-7}, which will be valid only when the SQUID cavities are driven at sufficiently low excitation powers, and where the junctions are still under a linear operation regime: 
\begin{equation}
L_{J}=\varphi_{0}(\frac{1}{I_{c}\cos{\phi}})\label{eq-8}.
\end{equation}

When operating in a linear regime (where $I<<I_{c}$), only the first term in Eq.~\ref{eq-7} plays a role as a Josephson inductance, and the total inductance of the cavity can be expressed as follows.
\begin{equation}
L_{tot}=L_{g}+\frac{\varphi_{0}}{I_{c}\mathrm{\cos}|\frac{\pi \Phi}{\Phi_{0}}|}\label{eq-9}
\end{equation}
Here, the first term on the right hand side of the equation is the geometric inductance of the resonator, the second term is the contribution of the Josephson junctions, and $\Phi$ is the applied flux through the SQUID loop.  Furthermore, from the expression of the resonance frequency of a flux tunable resonator $\omega(\Phi)=1/\sqrt{L_{l}(\Phi)C_{l}}$, one can also express this as
\begin{equation}
\omega(\Phi)=\frac{\omega_{0}}{\sqrt{1+\Gamma/\mathrm{\cos}(\frac{\Phi}{\Phi_{0}})}}\label{eq-10}
\end{equation}
where $\Gamma=\frac{L_{J}}{L_{J}+L{g}}$ is the SQUID inductance participation ratio and $\omega_{0}$ is the frequency of the cavity in the limit, $L_{J}\rightarrow 0$ .

\section{Estimation of critical current from normal state resistance}
\begin{figure}
\centering
\includegraphics[width=160mm]{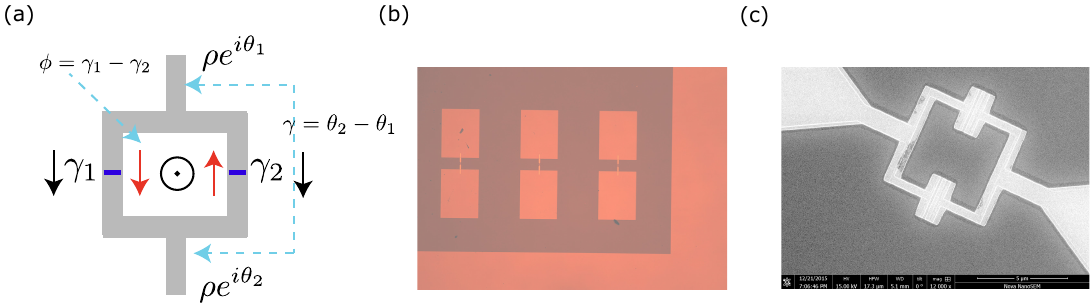}
\caption{
(a)figure shows a schematic diagram of a dc SQUID.  $\mathrm{\gamma_{1}}$ and $\mathrm{\gamma_{2}}$ are the superconducting phase of each junction indicated by black arrows. Red arrows indicate a circulating current. 
$\mathrm{\gamma}$ is the superconducting phase difference over the SQUID, and $\phi$ is the magnetic flux through the SQUID loop, which consists of the difference between the phases of each junction.  
(b) \textbf{Optical image of witness junctions.} There are 5 witness junctions on one chip. The contacting pads are made with MoRe, which is the same material as the other circuit elements.  Aluminium electrodes from evaporation of AlO$_{x}$ are connected galvanically to each MoRe contacting pads.
(c) \textbf{SEM image of a witness junction.} Each pair has a dc SQUID, the same as the measurement devices.}\label{FigS1}
\end{figure}

The measured samples contain reference Josephson junctions to estimate the critical current (we refer to them as witness junctions). Measuring the resistance of the witness junctions provides information about the evaporation, whether the junctions are shorted, open, or successful.  Inspection with a scanning electron microscope is helpful and tells us that the junction works well as long as two Al electrodes make good overlap, like the one shown in Fig.~\ref{FigS1}. 

A witness junction typically consists of two large contacting pads and a Josephson junction with metal electrodes that connect the two pads. Fig.~\ref{FigS1} shows witness junctions made for the device.  For each witness junction, the contacting pads are made of MoRe and the junctions and electrodes which connect the two pads are made of Al, separated by $\sim$~70~$\mu$m.  Since Al is a good electric conductor at room temperature compared with MoRe, the residual resistance of the connecting electrodes minimizes its contribution to the measured resistance.  The sample contains 5 identical witness junctions for each device: by comparing the resitance value of the 5 identical junction, one can estimate the junction yield of the batch.  Fig.~\ref{FigS1}~(b) is a zoomed image of a witness junction, which contains two Josephson junctions in a loop instead of a single junction.  The resistance of the junctions is measured using a probe station with a resistance box.  The critical current of a witness junction is estimated from the room temperature resistance $R_{n}$ and the superconducting gap of the Al. In the low-temperature regime, where $T<<T_{c}$, the Ambegaokar and Baratoff relation says
\begin{equation}
I_{0}=\frac{\pi\Delta}{2eR_{n}}\label{eq-12}
\end{equation} 
where $\Delta$, $e$, and $R_{n}$ are the superconducting gap of the Al, the electric charge, and the room temperature resistance over the junction.  The measured resistance of the witness junctions which contain two junctions in parallel is in the range between 500~$\sim$~1~K$\Omega$ using a `beeper box'.  On average, the room temperature resistance of witness junction is around 700~$\Omega$ on one SQUID, whose critical current is estimated to be $\sim$~490~nA using the equation above. The corresponding Josephson inductance would then be around 0.8~nH (Eq.~\ref{eq-8}). The value of the Josephson inductance found from the fit of the frequency dependence to the flux, is estimated to be around 0.35~nH. The room temperature resistance measured with the beeper box gives a ballpark estimate of $L_J$ and $I_c$ to within a factor of two.

\newpage
\section{Estimation of input power at the device}
We calculated the total attenuation at the sample stage in three different ways: (1) The input power is estimated by calculating the attenuation on the input line. (2) The input power is estimated from the background noise of the VNA and the gain of the amplifiers and the attenuation on the output line.  (3) The input power is estimated from the signal to noise ratio on the VNA.

First, we estimate the input power by calculating the total attenuation on the input line.  The microwave signal sent from a VNA is attenuated to reduce the thermal noise fluctuation.  The measurement setup is shown in Fig.~\ref{fig6-7}. The microwave signal sent from the output of the VNA goes into the input line of the fridge, which passes through multiple attenuators at each temperature stage.  For the input line, there are attenuators placed at each stage of the fridge, which add up to 47~dB.  
In addition, the microwave loss in the coaxial cables on the input line is estimated.  From the top flange to the 4~K stage, $SC-219/50-SCN-CN$ is used. The outer conductor is cupronickel with a diameter of 2.2~mm and a silver-plated cupronickel centre conductor.  The distance from top flange to the 50~K stage is 176~mm, and the distance from the 50~K stage to the 4~K stage is 270~mm.  
We estimated the attenuation between the top flange to the 4~K stage through the cupronickel coaxial cables to be 0.826~dB.  Below the 4~K stage, SC-086/50-SCN-CN is used.  The cable has a diameter of 0.86~mm, and is made of cupronickel with a silver-plated cupronickel inner conductor.  The attenuation in this cable is 3.2~dB/m at 4~K. 
The distance between the 4~K stage to the still flange is 240~mm, and from the still flange to the MC plate is 228~mm. 
The attenuation below the 4~K stage was estimated to be 1.498~dB. The total attenuation through the coaxial cable in the input line excluding the attenuators was estimated to be 2.32~dB.  At the output port of the VNA, extra attenuators to the amount of 60~dB were added.  
The coaxial cable between the output of the VNA and the top of the fridge has been estimated to be 3.54~dB at resonance frequency.  
The starting output power of the VNA was set to -30~dBm.  Therefore, the total power at the sample is around -142.86~dBm.
 
Second, the total attenuation was calculated from the attenuation and the gain in the output line.
For this estimation, the attenuation and the gain in the input line to the VNA, and the output power of the VNA.  The noise level of the VNA indicates the attenuation and gain, and the attenuation level at the device can be inversely calculated by subtracting the gain and adding the attenuation in the amplifier line. In one of the measurements, the measured noise level was -91.5 dBm.  At room temperature, two Meteq amplifiers were used for better visibility. Each amplifier has a gain of 28 dB. The gain of the HEMT amplifier is 37 dB around 4 to 8 GHz. From the cold temperature amplifier to the top of the fridge, we estimate the attenuation to be 1.8~dB, including the 1~dB attenuator. From the HEMT to the 10 mK stage, we ignored  cable losses because NbTi is superconducting at 3~K or below.  We considered each isolator's contribution to be a 0.2~dB loss. By starting at -30 dBm on VNA output power,  the power at the device is -182 dBm.

The third technique is to estimate the power at the device from the S/N of the measurement.
The two techniques mentioned previously lead to a discrepancy in the estimated values. The signal comes out of the amplifier line and goes into the input port of a VNA passing through two room temperature amplifiers. The fluctuating noise in the measurement is dominated by the Johnson noise of the HEMT amplifier, confirmed by observing the noise level with the HEMT amplifier on and off.  The thermal noise power at the input of the HEMT amplifier is determined by: 
\begin{equation}
P_{dBm}=10\log10(K_{B}T\times 1000)+10\log10(\Delta f)\label{eq-13}
\end{equation}
where $K_{B}$ is Boltzmann's constant, $T$ is the noise temperature of the HEMT, and $\Delta$f is the IF bandwidth of the VNA.
A factor of 1000 is used for conversion to dBm.  The noise temperature of the HEMT is 5.5~K\texttt{(LNF\_LNC1\_12A)}, and the noise power with an IFBW of 10~Hz gives -181.20~dBm, which corresponds to 0.195~nV in RMS voltage. The noise power is proportional to the attenuation between the HEMT and the input/output port of the device, the fluctuation of the measurement in the noise level, and some proportionality constant. 
\begin{equation}
\sigma_{V_{in}^{amp}}=\alpha A \sigma_{M}.\label{eq-14}
\end{equation}
The left-hand side of the equation is the power which goes into the HEMT amplifier. On the right-hand side, $\alpha$ is the attenuation between the HEMT amplifier and the output of the device, and $\sigma_{M}$ is the noise determined by taking the standard deviation of the measured voltage on the VNA.  The total attenuation from the HEMT amplifier to the input/output port of the device was determined to be 1~dB, considering each isolator to have an insertion loss of 0.2 dB. Here, $A$ is a proportionality constant so that
\begin{equation}
|V_{out}|=A\times M\label{eq-15}
\end{equation}
where $M$ is the measured transmission spectrum, in this case $|S_\mathrm{21}|$.  The voltage of the signal that goes into the HEMT is easily found by multiplying the attenuation factor by the input voltage to the HEMT.  The output voltage can be converted into power in dBm and photon number.
With this procedure, the estimated attenuation from the output of the VNA to the device is -155.77 dBm.
We believe the discrepancy in power at the device calculated from the first two procedures comes from estimating the attenuation or amplification gain with the wrong numbers, which most likely comes from loss in a connector/cable on the input and output line of the transmission line that has not been accounted for. 

\begin{table}
\caption{Important parameters of the devices}
\begin{center}
\begin{tabular}{| m{0.6cm} | m{6.0cm} | m{1.8cm}| m{1.8cm}| m{1.8cm}| m{1.8cm}|} 
\hline
\textbf{\#} & \textbf{Estimation Technique} & \textbf{P\textsubscript{in,min} (dBm)} & \textbf{N\textsubscript{ph,min}}&\textbf{P\textsubscript{in,max} (dBm)}&\textbf{N\textsubscript{ph,max}} \\
\hline
\ 1 & PNA output power + input line attenuation estimation& -142.86& 0.18 &-112.86&183.29\\ 
\hline
\ 2 &PNA input power + amplifier chain gain estimate& -182 & 0.00&-152&0.02 \\ 
\hline
\ 3 &PNA SNR + HEMT input noise estimate +attenuation loss between sample and HEMT estimate& -155.75& 0.01 &-125.75&9.41\\ 
\hline

\end{tabular}\label{tb6-2}
\end{center}

\end{table}

\begin{figure}
\centering
\includegraphics{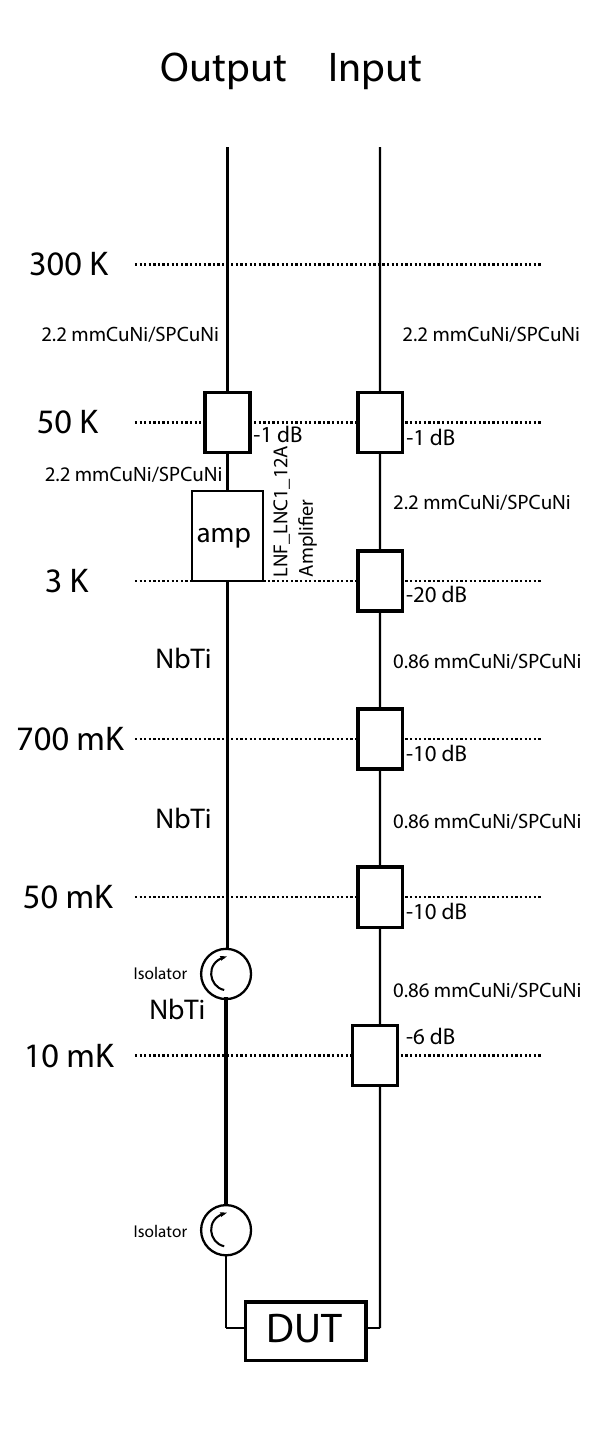}
\caption{\textbf{Measurement setup of one of the devices.}  }
\end{figure}\label{fig6-7}
\newpage
\section{Estimation of geometric capacitance and inductance of the device}

We determined the geometric inductance and capacitance of the device from a simulation. In the experiments, we measured a lumped element resonator and a hybrid SQUID cavity, in which both the resonators, made of MoRe, have the same geometry with only one difference: the hybrid SQUID cavity has a dc SQUID in the middle of the inductive element.  We first estimated the contribution of the kinetic inductance in the MoRe film. Fig.~\ref{fig6-9}~(b) is an optical image of a reference resonator which was made in the same batch as the hybrid device shown in Fig.~\ref{fig6-9}~(a). 
A Sonnet simulation of the same design as shown in Fig.~\ref{fig6-9}~(c) found that the resonance frequency of the reference resonator should be around 7.05~GHz. The observed resonance frequency of the mode is around 5.425~GHz.  The discrepancy between the resonance frequencies of the reference resonator in the measurement and in  the Sonnet simulation is due to the kinetic inductance of the 60~nm thick MoRe film.  In order to find the contribution of the kinetic inductance in terms of sheet resistance, we added sheet inductance until the resonance of $S_\mathrm{21}$ response in Sonnet  went down to 5.425~GHz. We found the corresponding sheet inductance to be 1.575~pH/sq.

Now the Josephson inductance of the SQUID can be found by comparing the resonance frequency of the reference resonator with 
the maximum frequency of the hybrid SQUID cavity. The difference in the frequencies is due to the Josephson inductance, which can be calculated in the simulation by introducing an ideal inductive element in the circuit until the frequency of the response matches the maximum frequency of the hybrid SQUID cavity.  From the simulation, the estimated Josephson inductance is 0.35~nH. 

Two unknown parameters which are still to be calculated are the geometric inductance and the geometric capacitance. The assumption we made here 
is that the geometric inductance and capacitance are constant with respect to the power or field strength, which is a reasonable assumption for the dynamic range of a dc 
SQUID.  Then we further increased the ideal inductive element, which corresponds to a frequency shift of the SQUID cavity either by Kerr non-linearity or tuning the flux through the SQUID loop.  The frequency shift in terms of the Josephson junction can be expressed as 
\begin{equation}
\omega_{0}=\frac{1}{2\pi\sqrt{(L_{g}+L_{j})C_{g}}}\label{eq-16}.
\end{equation}

\begin{figure}
\centering
\includegraphics{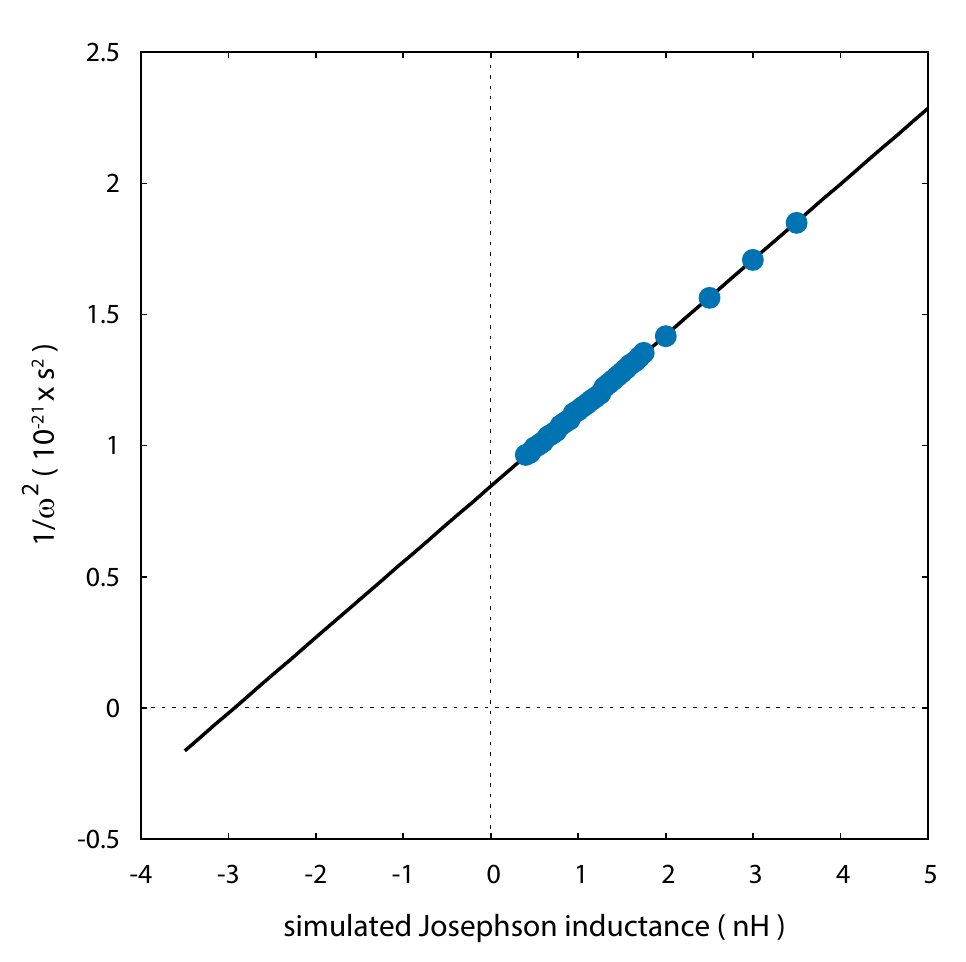}
\caption{Fit to $1/\omega^{2}(L_{J})$ as a function of ideal inductance. Blue points are numerical values from a simulation 
where the ideal inductance was varied point by point. The black line is a fit with Eq.~\ref{eq-16}. The lowest point of the inductance is 0.35~nH, which is the estimated Josephson inductance at integer flux quantum. The slope crosses of the $x$-axis on $\sim-2.9~nH$, which corresponds to the geometric inductance of the resonator with a negative sign. }\label{fig6-8}
\end{figure}

From the equation above, both unknown parameters can be determined simultaneously. 
Fig.~\ref{fig6-8}  shows the fit using Eq.~\ref{eq-13}.  The blue points are simulated points whose $y$-values are the square of the resonance frequency and whose $x$-values are the values of the ideal inductor. The black line is the fit to the points.  The fit works nicely with geometric inductance, L\textsubscript{g}=2.93nH 
and geometric capacitance, C\textsubscript{g}=288~fF. Josephson inductance participation ratio 0.11, which is in a good agreement with the value estimated from the curve of frequency under flux which is shown in Fig.~\ref{fig3}~(b). 
 
\begin{figure}
\centering
\includegraphics[width=150.0mm]{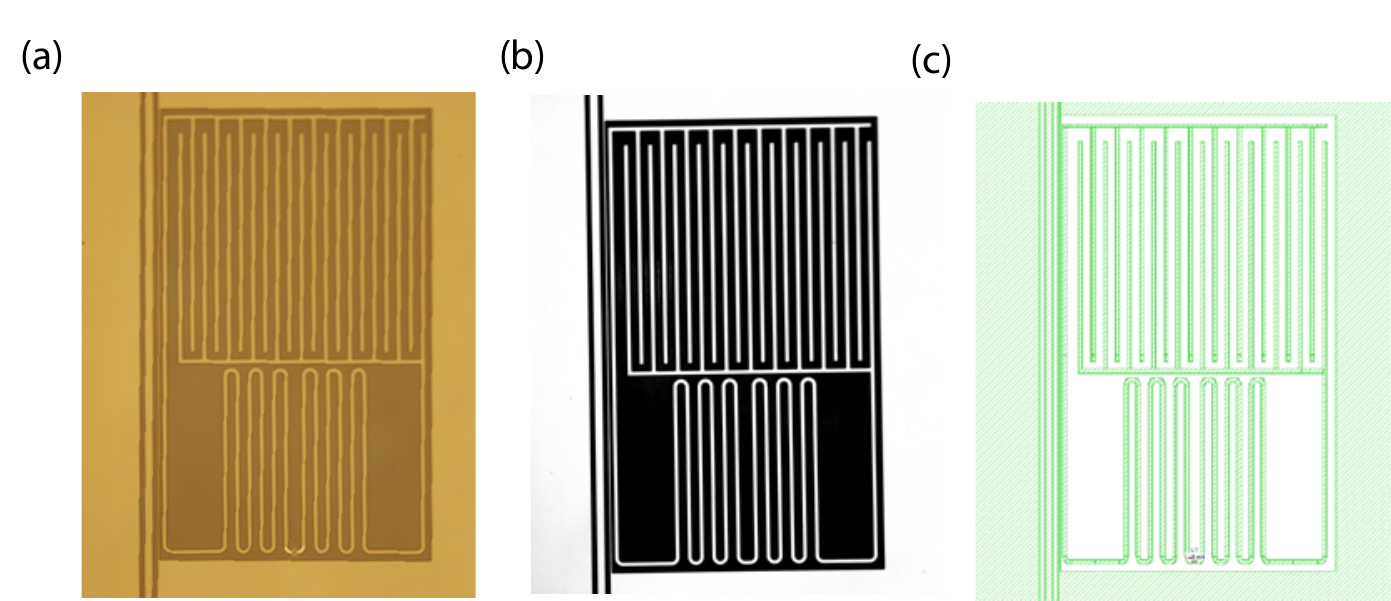}
\caption{(a) Optical microscope image of a hybrid resonator. (b) Optical microscope image of a reference lumped element resonator having the same geometry as the hybrid device shown left. (c) Screen shot image of the design used for estimating resonance frequency of the device.}\label{fig6-9}
\end{figure}

\section{Asymmetric Response fitting}
\begin{figure}
\centering
\includegraphics[width=100.0mm]{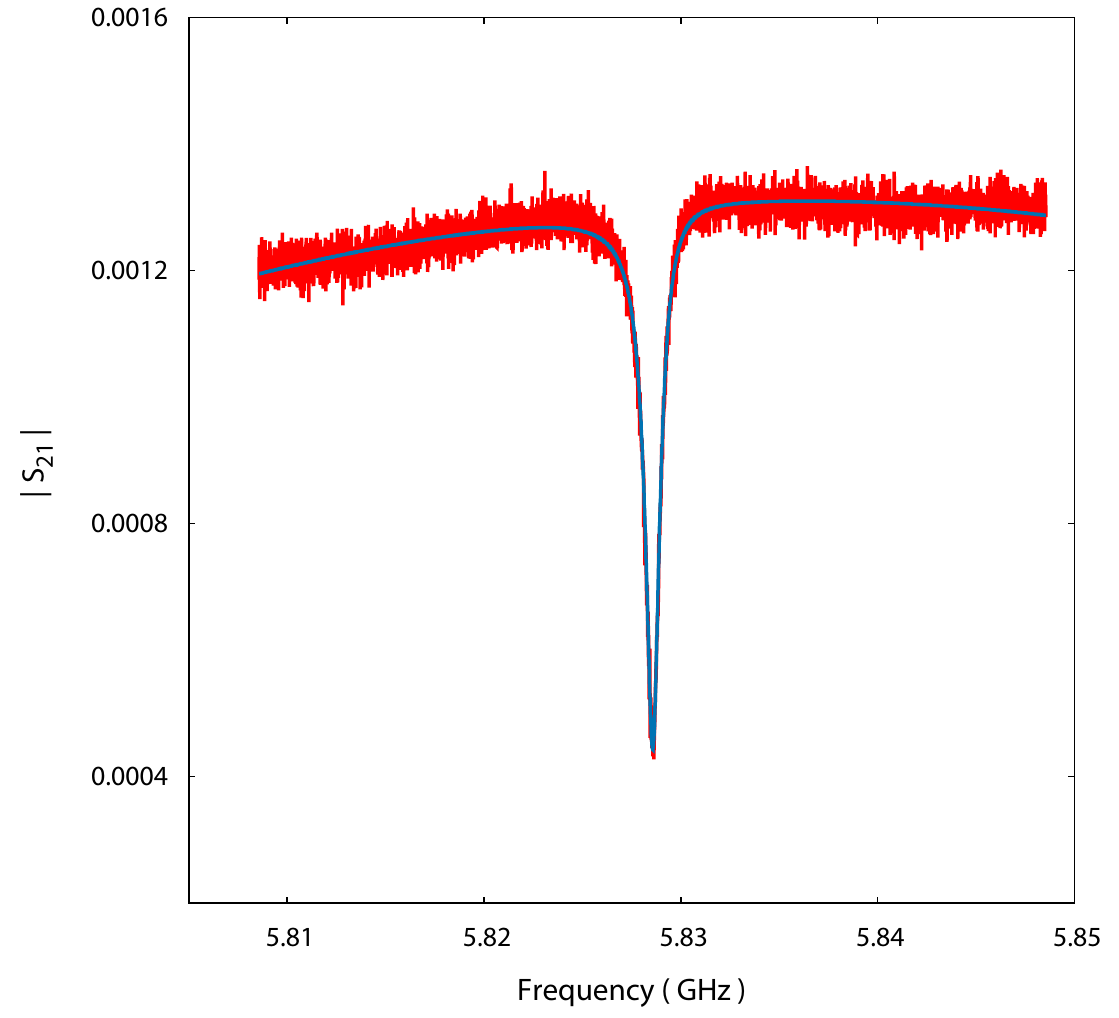}
\caption{One of the measured responses of a linear CPW resonator(\textbf{REF\_CPW2}) and fit.  For this fit, the fitting parameters were $f_{0}$=5.83~GHz, $Q\textsubscript{int}$= 15.08$\times \mathrm{10^{3}}$, Q\textsubscript{ext}=7.72$ \times \mathrm{10^{3}}$, a= -4.86, b=1.67e-09, c=-1.43e-19, a'=2.21e+03, b'=-3.80e-07, and $\theta$=-0.09$\degree$.}\label{fig6-10}
\end{figure}

When we measure the quality factor of a microwave resonator, we get two quality factors, the internal and the external quality factor. In measurements of superconducting resonators, small signal reflections or non-negligible impedances could give rise to an asymmetric Lorentzian response.  
The transmission spectrum of a side coupled resonator, also called the notch type geometry, is given by\cite{probst_2015}.
\begin{equation}
S_\mathrm{21}=ae^{i\alpha}e^{-2\pi i f \tau}[1-\frac{Q_{loaded}/{|Q_{ext}|}e^{i\theta}}{1+2iQ_{loaded}(f/f_{r}-1)}]\label{eq-18}.
\end{equation}
The prefactor describes non-ideal events in the measurement line, such as attenuation, impedance mismatch, and cable delay. 
Inside the brackets is the response of an ideal side coupled cavity. The term is a complex Lorentzian function which is subtracted from unity. 
The resonance response has the relation such that $1/Q_{loaded}=1/{Re}~[Q_{ext}]+1/Q_{int}$, where the external quality factor, $Q_{ext}$, is a complex value such that $Q_{ext}=Q_{ext}e^{-i\theta}$.
The ideal resonance response term can be rewritten in terms of the cavity decay rates:
\begin{equation}
S_\mathrm{21}=1-\frac{\kappa_{ext}e^{i\theta}}{\kappa_{loaded}+2i\delta\omega}\label{eq-19}
\end{equation}
where $\kappa_{ext}$, $\kappa_{loaded}$, and $\delta\omega$ are the external decay rate, total decay rate, and detuning of the drive tone from the resonance frequency. Here $\theta$ plays a role to compensate for the asymmetry in the resonance due to any impedance mismatch in the measurement line. 
The formula below was used to plot the cavity response, including environmental effects. 
\begin{equation}
S_\mathrm{21}=Ae^{i(a^{'}+b^{'}\omega)}[1-\frac{\kappa_{ext}e^{i\theta}}{\kappa_{loaded}+2i\delta\omega}]\label{eq-20}
\end{equation}
The prefactor of Eq.~\ref{eq-20} is equivalent to that of Eq.~\ref{eq-18}, except we assume a small fluctuation of the background which depends on the frequency, 
$A=a+b\omega+c\omega^{2}$, where $a$, $b$, and $c$ are fitting parameters of background of zeroth, first, and second order, respectively. 
$a'$ and $b'$ play the roles of phase shift and electrical length of the meter-long coaxial cable. Any impedance mismatch of the measurement line 
causes a rotation of the complex-resonance circle in the $I-Q$ plane, and $\theta$ takes into account the accumulated rotation due to an impedance mismatch.
Here, the fitting is done with the amplitude of $S_\mathrm{21}$.  Fig.~\ref{fig6-3}~(c) is fitted with  
Eq.~\ref{eq-19}.  The transmission spectrum goes above unity on the right side of the plot, which is an artifact resulting from the accumulated phase $
\theta$ from any impedance mismatch inside the connecting cables or connectors or wirebonds.
Fig.~\ref{fig6-10} shows a Lorentzian response of one of the linear reference CPW resonators.  The response is fitted with the function above with the parameters below.

\section{Resonance frequencies of the circuit}
\begin{figure}
\centering
\includegraphics[width=100.0mm]{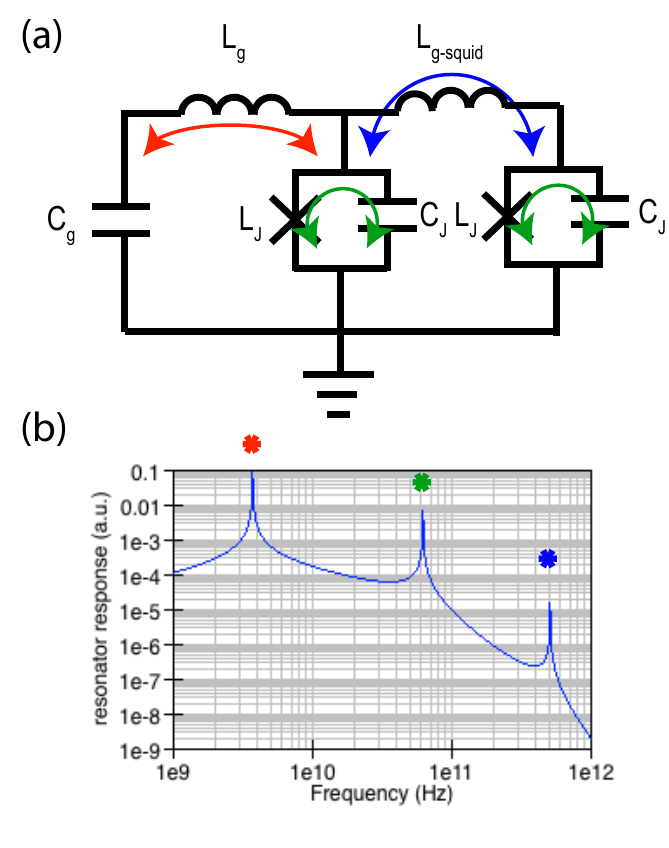}
\caption{\textbf{QUCS simulation of possible resonances in the circuit.} (a) Circuit representation of our hybrid device. There are three possible resonances: (1) Resonance frequency of SQUID cavity is indicated by red arrow. (2) Plasma frequency of two Josephson junction is represented by green arrows. (3) Circulating current mode is indicated by blue arrow.  This circulating current mode satisfies the condition that $L_{J}>>L_{g-squid}$. The resonance mode is created by the geometric inductance of the SQUID and the capacitance of the Josephson junctions.  (b) QUCS simulation of an equivalent circuit to our device.  Here we estimate resonance frequencies with parameters close to the device values, where L\textsubscript{J}= 350~pH,  C\textsubscript{J}=20fF, L\textsubscript{g-squid}=20~pH, L\textsubscript{g}=2.9~nH, and C\textsubscript{g}=0.6~pF.  Three resonances are observed. The lowest resonance corresponds to the SQUID cavity and is indicated by a red asterisk. The second highest mode is around 60~GHz. This mode is the mode of the plasma freuency of two Josephson junctions coupled via the small SQUID loop inductance, L\textsubscript{g-squid}. The highest one is the circulating mode, whose resonance lies around 400~GHz. The circulating current mode has a phase dependence on time, but this is neglected since the dynamics of the highest mode moves fast compared with the lowest mode, which is our interest.  }\label{fig6-11}
\end{figure}

The resonance frequencies of a SQUID cavity was simulated in the Quite Universal Circuit Simulator (QUCS).  Fig.~\ref{fig6-11}~(a) is a circuit representation of one of our devices, where $L_{g}$ and $C_{g}$ are the geometric inductance and capacitance, respectively. In the device, a meander strip forms a geometric inductance, $L_{g}$. The interdigitated capacitor in Fig.~\ref{fig6-9} is equivalent to $C_{g}$ in Fig.~\ref{fig6-11}~(a). $L_{J}$, $C_{J}$, and $L_{g-squid}$ are the Josephson inductance of the junction, the parallel plate capacitance of the junction, and the geometric inductance of the SQUID loop.  In this circuit, there are three resonances, represented in three different colours in Fig.~\ref{fig6-11}~(a). The red arrow represents a mode consisting of the interdigitated capacitor, a meander inductor, and a SQUID inductance. We use this mode for the measurements in the main text.  The green arrow represents a mode of two coupled Josephson junctions via a small SQUID loop inductance. From the design of our Josephson junction and the size of the SQUID loop, this mode should have a much higher frequency than the one represented by the red arrow. The blue arrow indicates a mode between the geometric inductance of the SQUID loop and the parallel plate capacitance from two Josephson junctions.  Compared with the mode represented by the green arrow, this mode should have a much higher frequency, since $L_{J}>>L_{g-squid}$. 

Fig.~\ref{fig6-11}~(b) shows a simulation of the voltage response of the circuit represented in Fig.~\ref{fig6-11}~(a) with parameters close to the experimental values.   Three resonances are observed. The lowest mode corresponds to the red arrow in Fig.~\ref{fig6-11}~(a).  The middle mode corresponds to the green arrow in the figure above. The highest mode corresponds to the blue arrow.  In the measurement, the two higher modes cannot be observed in the measurements because the measurement bandwidth is limited to 4$\sim$8~GHz.  The mode which is associated with the circulating current of the SQUID (oscillating $\phi$) is the highest mode, whose resonance frequency lies around 400~GHz. The time derivative of the circulating current mode becomes approximately flat over many oscillations during the time scale corresponding to a frequency of the 6~GHz resonance frequency. 

\newpage

\section{Contamination from resist sidewalls during liftoff process}

\begin{figure}
\centering
\includegraphics[width=80.0mm]{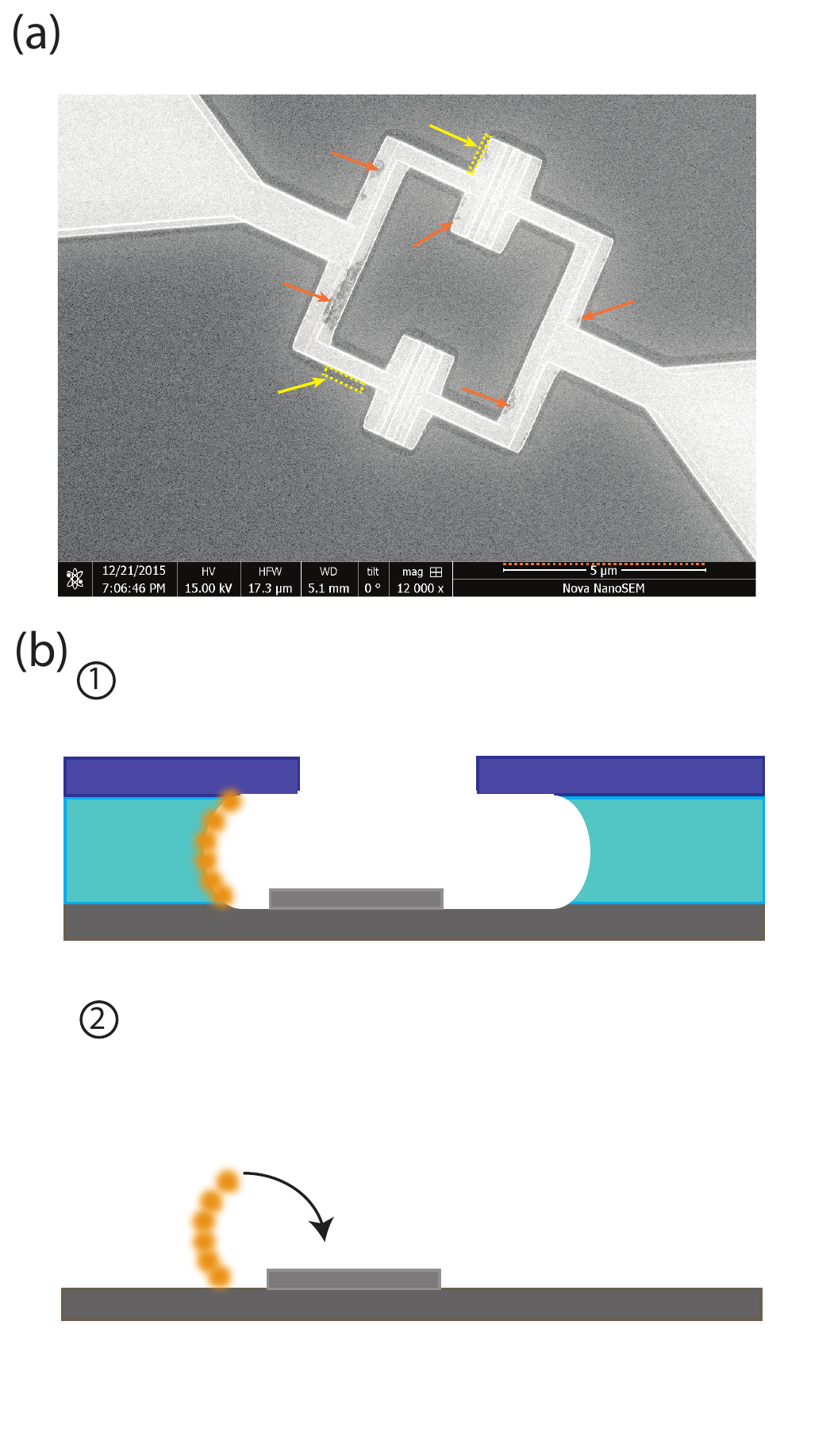}
\caption{(a) SEM image of a dc SQUID. The yellow arrows indicate a darker colour compared with the area away from the dc SQUID.  These darker regions have seemingly less resist residue in those regions, and indicate the size of the bottom layer undercut.  During oxygen plasma or BHF etching or a combination of both, resist residue is removed and produces this different colour away from the SQUID.  There are dark spots on top of the SQUID, which are indicated with orange arrows.  They are probably the residue of the electron beam resist.  This resist residue is known as the black veil of death \cite{slichter_quantum_2011}. (b) Possible formation of black veil of death: $\mathrm{\textcircled{1}}$. The chemical reaction of this sidewall occurs during evaporation. $\mathrm{\textcircled{2}}$ Free standing sidewall falls over during lift off. }\label{fig6-12}

\end{figure}

Fig.~\ref{fig6-12}~(a) shows a SEM image of a dc SQUID after lift off. There are several parts which appear darker on the SQUID loop. These parts are known as the black veil of death \cite{slichter_2011}, which is a resist residue sidewalls of possibly the resist stack flipping over on top of the device during the lift off of the Josephson junctions.  This could be a result of a chemical reaction of the electron beam resist during the evaporation, and these contaminations falling over on top of the device as shown in Fig.~\ref{fig6-12}~(b).
 
 For the hybrid devices of our measurements, the internal quality factors are not limited by dielectric loss due to the carbon contamination of resist residue in the Josephson junctions because their coherences are comparable to those of the reference devices.  However, for the devices made with the single step lift off, the resist residue might not be negligible, and post oxygen ashing with oxygen plasma should be considered to reduce the dielectric loss.  

Post cleaning with oxygen plasma is known to remove the black veil of death.  We believe that these observations indicated it is likely beneficial to perform a descum plasma cleaning step after lift off.  
 
\end{document}